\documentclass[journal]{new-aiaa}
\usepackage[utf8]{inputenc}
\usepackage{textcomp}

\usepackage{graphicx}
\usepackage{amsmath}
\usepackage[version=4]{mhchem}
\usepackage{siunitx}
\usepackage{longtable,tabularx}
\usepackage{makecell} 
\usepackage{multirow} 
\usepackage{subfigure}
\title{ Gain-Scheduled Passive Fault-Tolerant Control Design for Dual-System UAV Transition Flight }

\author{Junfeng Cai \footnote{Ph.D. Candidate, Department of Aerospace Science and Technology, via La Masa 34.} and Marco Lovera\footnote{Full Professor, Department of Aerospace Science and Technology, via La Masa 34.}}
\affil{Politecnico di Milano, Milan, 20151, Italy}

\begin{document}

\maketitle

\begin{abstract}
Dual-system UAVs with vertical take-off and landing capabilities have become increasingly popular in recent years. As a safety-critical system, it is important that a dual-system UAV can maintain safe flight after faults/failures occur. This paper proposes a gain-scheduled passive fault-tolerant control (PFTC) method for the transition flight of dual-system UAVs. In this novel FTC design method, the model uncertainties arising from the loss of control effectiveness caused by actuator faults/failures, for the first time, are treated as model input uncertainty, allowing us to use multiplicative uncertainty descriptions to represent it. The advantages of the proposed method consist in significantly reducing the number of design points, thereby simplifying the control synthesis process and improving the efficiency of designing the FTC system for dual-system UAV transition flight compared with the existing FTC design methods. As a general method, it can be applied to the design of FTC systems with multiple uncertain parameters and multiple channels. The developed passive FTC system is validated on a nonlinear six-degree-of-freedom simulator. The simulation results demonstrate that the gain-scheduled structured $H_{\infty}$ (GS SHIF) PFTC system provides superior fault tolerance performance compared with the LQR and structured $H_{\infty}$ control systems, thereby showcasing the effectiveness and the advantages of the proposed GS SHIF PFTC approach.
\end{abstract}

\section{Introduction}
Unmanned aerial vehicles (UAVs) have been widely applied to many fields, including the military and civil applications. Hybrid configuration UAVs have particular advantages over the traditional ones. Compared to fixed-wing UAVs, they can vertically take off or land without runways. Compared to multicopter UAVs, they have lower energy consumption during cruise flight. The Dual-system UAV as a type of hybrid UAVs has a relatively simple mechanical structure and smooth attitude variations during transition. Due to these advantages, in recent years dual-system UAVs have been applied to logistics and urban air mobility that mainly operate over urban areas. Therefore, the flight safety and reliability of dual-system UAVs are critical. If the control systems of dual-system UAVs can maintain safe flight even if there is possible occurrence of faults/failures in the systems, then the flight safety and reliability of dual-system UAVs can be improved, namely the development of the fault tolerant capabilities of the control systems. Therefore, the investigation of the fault-tolerant control (FTC) design of dual-system UAVs is of significance. 

Fault-tolerant flight control is aimed at designing flight control systems that can maintain safe flight even if there is occurrence of faults/failures. FTC can be divided into passive FTC (PFTC) and active FTC (AFTC). PFTC consists in designing control systems that are robust to possible faults/failures in systems. After fault occurrence, there is no need of controller reconfiguration and fault information. In contrast, AFTC eliminates the adverse effects caused by the faults/failures by changing controller structures or parameters, where the fault information including fault locations and magnitudes is necessary. Therefore, an AFTC system typically includes a fault detection and diagnosis (FDD) module that provides fault information for controller reconfiguration.  

Hybrid UAVs as over-actuated systems provide the possibility of developing the fault-tolerant capabilities of control systems. As safety-critical systems, the FTC of Hybrid UAVs has drawn the attention of researchers.  With this regard, there have been some interesting results. In \cite{1}, an adaptive sliding mode AFTC method was proposed for an hybrid vertical take-off and landing canard rotor/wing unmanned aerial vehicle (UAV) to handle the actuator faults and model uncertainties. In \cite{2}, a sliding mode AFTC system was developed to handle the actuator faults/failures for an octoplane UAV in the transition mode. In \cite{1,2}, the sliding mode control (SMC)-based FTC approach can effectively deal with the actuator faults/failures of hybrid UAVs. However, first, SMC is a discontinuous control strategy, which usually results in control chattering in the states of UAVs. Second, even though in \cite{1} the developed adaptive SMC FTC system can deal with the actuator faults and unmodeled dynamics simultaneously, this method is not able to cope with the varying dynamics of dual-system UAVs during transition flight that is parametric model uncertainty. In addition, in both \cite{1} and \cite{2}, fault information is needed and the FTC performance is affected by the fault information error and delay. In \cite{3}, a novel incremental adaptive sliding mode control was proposed to handle the aerodynamic model uncertainties and rotor faults in the multicopter mode. Where the FTC design of transition mode is not involved, and also the fault information of the actuators is needed for the developed FTC system. In \cite{4,5,6}, the FTC of a dual-system UAV in the fixed-wing flight mode was investigated using active fault-tolerant control methods, in which the robustness of the developed FTC system to model uncertainties is not discussed and the fault detection and diagnosis module is required.

The structured $H_{\infty}$ approach has been applied to the FTC area in recent years. The main structured $H_{\infty}$-based FTC method is referred to as self-scheduled FTC. This methodology has already been used for the FTC of multicopter and fixed-wing aircraft. In \cite{7}, a self-scheduled AFTC system for a quadrotor UAV involving both fault detection and diagnosis (FDD) and FTC is presented. The AFTC system resorts to the gain-scheduling technique with controllers synthesized by the structured $H_{\infty}$ method at discrete points. In \cite{8}, a universal self-scheduled AFTC method is proposed, and the effectiveness and the robustness of this method are demonstrated by the high fidelity simulations and experiments. In \cite{9}, robust self-scheduled fault tolerant control laws are designed for a multicopter UAV, in which two different variants for self-scheduled FTC are compared. The experimental results show the validity of the developed FTC system. In \cite{10}, a self-scheduled fault-tolerant controller is designed using structured $H_{\infty}$ control for the lateral/directional motion control of a fixed-wing aircraft. The aircraft-in-the-loop simulations show the good behavior of the aircraft under the defined failure scenarios. In these existing research, the structured $H_{\infty}$-based self-scheduled FTC method has successful applications. However, it is subject to some limitations and drawbacks. Provided that there is parametric model uncertainty when designing FTC systems using the self-scheduled structured $H_{\infty}$ approach, the scheduling variables, including the uncertain parameters in the control-oriented models and the loss of control effectiveness, will lead to a large number of design points and at each design point structured $H_{\infty}$ control synthesis needs to be done, resulting in a significantly increased design workload. Moreover, the self-scheduled structured $H_{\infty}$ FTC as an AFTC method heavily relies on fault information to adaptively adjust the controller parameters (i.e., gain-scheduling), which means it is necessary to design FDD modules additionally.

In this paper, a gain-scheduled FTC method for dual-system UAV transition flight is proposed. Firstly, the control-oriented model considering the varying dynamics of transition flight as well as the loss of control effectiveness is presented. The actuator redundancy management by control allocation is introduced. Then, based on the control-oriented model, structured $H_{\infty}$ control synthesis using a multi-model approach is carried out for all the design points, in which multiplicative uncertainty descriptions are employed to represent the model uncertainties resulting from both the aerodynamic coefficient uncertainties caused by airspeed variation and the loss of control effectiveness induced by actuator faults/failures. These uncertainty descriptions are further integrated into the structured $H_{\infty}$ synthesis. Next, the robust stability and performance of the synthesized controllers at all the design points are examined, showing that the robust stability and performance of the resulting controllers at these design points are ensured. Eventually, the developed passive FTC system is validated on a nonlinear six-degree-of-freedom simulator. The simulation results of the gain-scheduled structured $H_{\infty}$ (GS SHIF) FTC are compared to those of LQR and structured $H_{\infty}$ control, demonstrating the effectiveness and the advantages of the proposed GS SHIF FTC approach.

The contributions of this paper can be summarized as following aspects:
\begin{enumerate}
	\item A novel gain-scheduled FTC design method is proposed for the dual-system UAV transition flight. In this method, the model uncertainty arising from the loss of control effectiveness caused by possible actuator faults/failures is equivalent to the model input uncertainty, which allows us to deal with not only the model uncertainty resulting from actuator faults/failures but also multiple parametric uncertainties. Therefore, the robustness of the developed FTC systems to model uncertainties can be enhanced.
	\item Provided that there are multiple uncertain parameters in the model apart from the loss of control effectiveness, in this case, the existing self-scheduled FTC design approach will lead to a large number of design points. The GS SHIF FTC method is able to reduce the number of design points by representing the model uncertainty with multiplicative uncertainty descriptions, which simplifies the control synthesis process and improves the efficiency of designing the FTC system for transition flight of dual-system UAVs.
	\item The developed GS SHIF FTC system is capable to realize the transition flight of the dual-system UAV under faulty conditions. Additionally, it can provide better command tracking performance compared to LQR and structured $H_{\infty}$ control.
	\item The proposed GS SHIF FTC system does not rely on the fault information, which avoids the design of the FDD module. At the same time, the FTC performance is not affected by the fault information error and delay.
\end{enumerate}

The rest of this paper is organized as follows. Section \ref{Sec.2} introduces the dual-system UAV and the control-oriented model that will be used for the subsequent control design. Section \ref{Sec.3} presents the fault-tolerant control design based on the multi-model approach using the structured $H_{\infty}$ control method. Section \ref{Sec.4} describes the analyses of the robust stability and performance of the resulting controllers at the discrete design points. In section \ref{Sec.5}, the simulation results are reported and analyzed. Section \ref{Sec.6} concludes the paper.       
                 
\section{Problem Formulation}\label{Sec.2}

\subsection{The Dual-system VTOL UAV}
\begin{figure}
	\centering
	\includegraphics[width=0.7\linewidth]{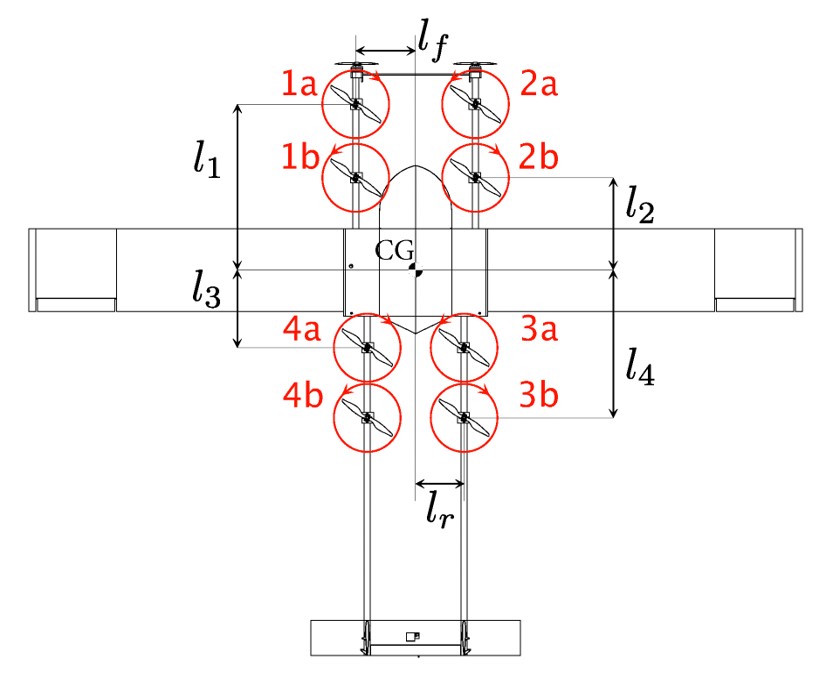}
	\caption{The top view of the dual-system UAV}
	\label{Figure1}
\end{figure}

The dual-system UAV possesses a hybrid configuration (see Fig. \ref{Figure1}), which includes fuselage, wings, eight vertical rotors, two horizontal rotors, ailerons, elevator, and rudders. Where $l_{1},\ l_{2},\ l_{3},\ l_{4},\ l_{r},\ l_{f}$ denote the rotor position dimensions. $1a,\ 1b,\ 2a,\ 2b,\ 3a,\ 3b,\ 4a,\ 4b$ represent different rotors, respectively. $CG$ represents the center of gravity of the UAV.

A complete flight mission of the dual-system UAV from takeoff to landing consists of several flight phases corresponding to different flight modes, i.e., multicopter mode, fixed-wing mode and transition mode. The transition mode is an intermediate mode from the multicopter mode to the fixed-wing mode. During the transition mode, all the actuators are activated. This paper mainly focuses on the investigation of fault-tolerant control of the transition mode of the dual-system UAV by assuming possible occurrence of actuator faults/failures. 

\subsection{Control-oriented Model}
The dynamics of the dual-system UAV during transition flight are complicated, during which all the actuators are working and the airspeed is increasing. Since the aerodynamic coefficients for the fixed-wing flight are varying with the airspeed, this poses a great challenge to the modeling of the transition flight dynamics. These varying aerodynamic coefficients with the airspeed lead to the varying transition dynamics of the UAV that are airspeed-dependent. 

To characterize the varying transition dynamics, in \cite{11}, linearized airspeed-dependent attitude dynamics models of dual-system UAV transition flight are derived. It is assumed that the transition flight is constant altitude steady level flight. Since the transition flight usually lasts only a few seconds. The horizontal position control is not the main concern. Therefore, we basically focus on the altitude dynamics and attitude dynamics. Further, if we also consider possible occurrence of actuator faults/failures, the loss of control forces and moments effectiveness can be introduced into the linearized attitude dynamics models. 

With reference to \cite{11}, the attitude dynamics models for the FTC design of the transition fight is thereby given as follows.

\subsubsection{Roll Dynamics}   
The transfer function from the roll rate $p$ to the rolling moment $M_{a_{x}}$ is given by:    
\begin{equation}\label{equ.1}
	G_{p}(s;V)=\frac{p(s)}{M_{a_{x}}(s)}=\frac{1-\gamma_{L}}{J_{x}s - M_{x}^{p}(V)}
\end{equation}
where $V$ is the airspeed. $J_{x}$ is the moment of inertia about $x$ axis. $\gamma_{L}$ is the loss of control effectiveness of the rolling moment. $M_{x}^{p}(V) = \bar{q}SbC_{L_{p}}(V)=\bar{q}SbC_{l_{p}}(V)\frac{b}{2V}$, denoting the derivative of the rolling moment with respect to the roll rate. $\bar{q}=\frac{1}{2}\rho V^{2}$ denotes the dynamic pressure. $S$ denotes the wing area. $b$ denotes the wing span. $C_{L_{p}}$ is the aerodynamic derivative of the rolling moment with respect to the roll rate. $C_{l_{p}}$ is the corresponding dimensionless quantity.

\subsubsection{Pitch Dynamics}
The transfer function from the pitch rate $q$ to the pitching moment $M_{a_{y}}$ is expressed as:
\begin{equation}\label{equ.2}
	G_{q}(s;V) = \frac{q(s)}{M_{a_{y}}(s)} = \frac{(1-\gamma_{M})s}{J_{y}s^{2} - M_{y}^{q}(V)s - M_{y}^{\alpha}(V)}
\end{equation}
where $J_{y}$ is the moment of inertia about $y$ axis. $\gamma_{M}$ is the loss of control effectiveness of the pitching moment. $M_{y}^{\alpha}(V) = \bar{q}S\bar{c}C_{M_{\alpha}}(V)$ represents the derivative of the pitching moment with respect to the angle of attack. $C_{M_{\alpha}}$ represents the aerodynamic derivative of the pitching moment with respect to the angle of attack. $\bar{c}$ represents the mean aerodynamic chord. $M_{y}^{q}(V) = \bar{q}S\bar{c}C_{M_{q}}(V) = \bar{q}S\bar{c}C_{m_{q}}(V)\frac{\bar{c}}{2V}$ represents the derivative of the pitching moment with respect to the pitch rate. $C_{M_{q}}$ represents the aerodynamic derivative of the pitching moment with respect to the pitch rate. $C_{m_{q}}$ is the corresponding dimensionless quantity.    

\subsubsection{Yaw Dynamics}
For the yaw dynamics model, the transfer function from the yaw rate $r$ to the yawing moment $M_{a_{z}}$ is given by:
\begin{equation}\label{equ.3}
	G_{r}(s;V)=\frac{r(s)}{M_{a_{z}}(s)}=\frac{(1-\gamma_{N})s}{J_{z}s^{2} - M_{z}^{r}(V)s-M_{z}^{\beta}(V)}
\end{equation}
where $J_{y}$ is the moment of inertia about $z$ axis. $\gamma_{N}$ is the loss of control effectiveness of the yawing moment. $M_{z}^{r}(V) = \bar{q}SbC_{N_{r}}(V) = \bar{q}SbC_{n_{r}}(V)\frac{b}{2V}$ represents the derivative of the yawing moment with respect to the yaw rate, $C_{N_{r}}$ represents the aerodynamic derivative of the yawing moment with respect to the yaw rate, $C_{n_{r}}$ is the corresponding dimensionless quantity. $M_{z}^{\beta}(V) = -\bar{q}SbC_{N_{\beta}}(V)$ represents the derivative of the yawing moment with respect to the sideslip angle, $C_{N_{\beta}}$ is the aerodynamic derivative of the yawing moment with respect to the sideslip angle. 

It is noticed that all the linearized attitude dynamics models are related to the airspeed which is a variable during transition flight. Since the aerodynamic coefficients obtained by a computational fluid dynamics (CFD) analysis are calculated with the airspeed as a scheduling parameter, the corresponding attitude dynamics models are functions of the airspeed as well, which is referred to as \textit{airspeed-dependent linearized attitude dynamics models}.

Usually, the loss of control effectiveness of forces and moments $\gamma_{T}, \gamma_{L}, \gamma_{M}, \gamma_{N}$ in the above equations are within the range of $[0,1]$. 0 means there is no faults/failures in actuators, 1 denotes complete actuator failures. In this paper, since we do not discuss the case of complete failures of all the actuators and at most four vertical rotors' complete failures are allowed, the ranges of the loss of control effectiveness $\gamma_{T}, \gamma_{L}, \gamma_{M}, \gamma_{N}$ are thereby chosen as $[0,0.6]$.      

\subsection{Actuator Redundancy Management}
As already described, the dual-system UAV is a typical over-actuated system. The redundant actuators are managed by an independent module whose function is to calculate the actuator commands according to the outputs of the control law, namely control allocation. In this paper, the actuator dynamics and the formulation of the control allocation problem are the same as those in \cite{12}. 

\section{Fault-tolerant Control Design Using a Multi-model Approach}\label{Sec.3}
In the previous section, the linearized airspeed-dependent attitude dynamics models are described. Given that the dynamics models during transition are varying with the airspeed and the loss of control effectiveness, it is impossible to use a single model for the FTC design for the transition flight. In this context, a multi-model approach is applied. That is to take discrete values of the airspeed. For each discrete airspeed value, structured $H_{\infty}$ control synthesis is carried out, leading to a set of discrete controllers at the design points. Based on the synthesized controllers, the GS SHIF FTC system is realized resorting to the gain-scheduling technique using piecewise linear interpolation method. 
   
\subsubsection{Mixed Sensitivity Optimization Based on Structured $H_{\infty}$}\label{Sec.3.1}
To deal with the model uncertainties induced by the variation of both aerodynamic coefficients and the loss of control effectiveness, a robust control approach, i.e., structured $H_{\infty}$, is employed for the control design. This structured $H_{\infty}$ method can solve $H_{\infty}$ optimization problems while maintaining fixed control structures and constraints on controller orders. Given that transition dynamics are fundamentally dependent on airspeed, using a single model for the entire transition control design is challenging. The structured $H_{\infty}$ approach employed here can adapt to varying dynamics and the fixed-structure controller, providing greater efficiency in handling multiple design points.

\begin{figure}
	\centering
	\includegraphics[width=0.7\linewidth]{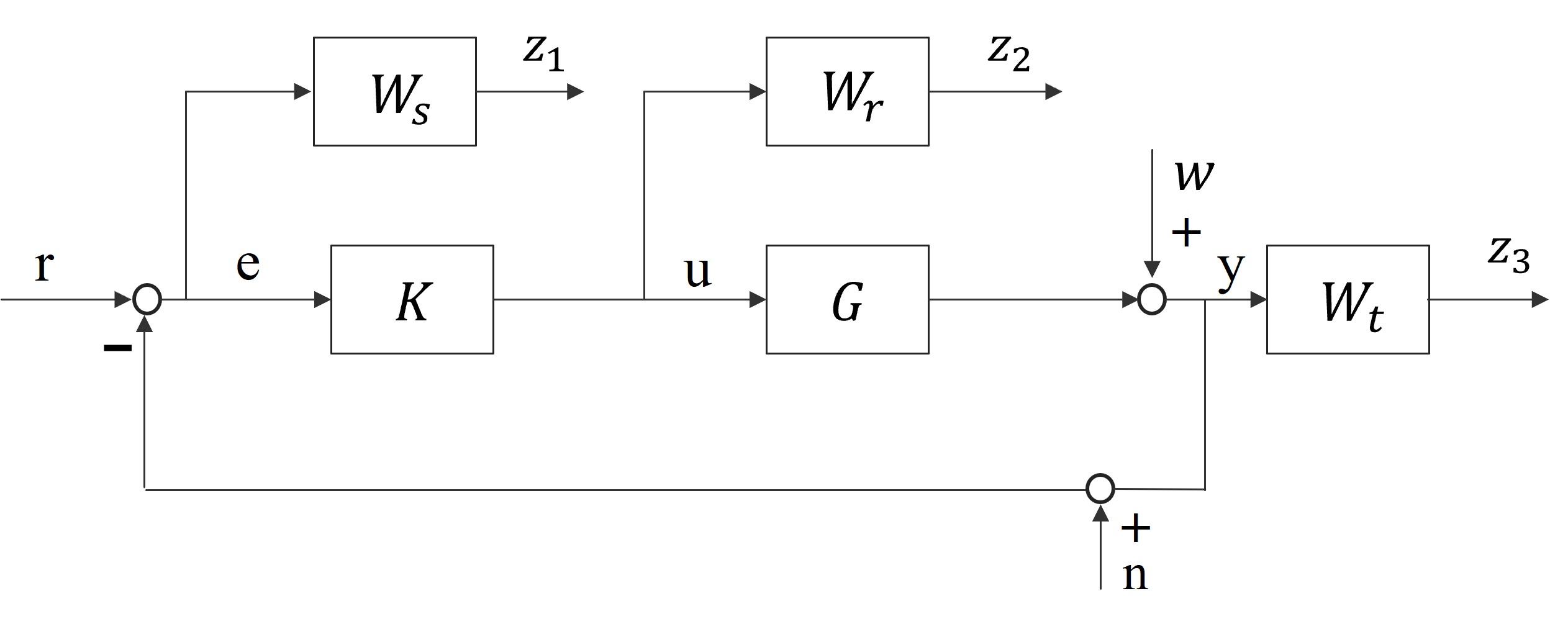}
	\caption{The general tuning block diagram of structured $H_{\infty}$}
	\label{Figure2}
\end{figure}

Structured $H_{\infty}$ solves the control parameters of the predefined control structures by optimizing the $H_{\infty}$ norms of the closed-loop transfer functions. The specific closed-loop transfer functions to be optimized include the sensitivity function $S(s)$, complementary sensitivity function $T(s)$ and control sensitivity function $R(s)$, respectively, as defined below:
\begin{equation}\label{equ.4}
	\begin{aligned}
		S(s) &= (I + G(s)K(s))^{-1}\\
		T(s) &= (I + G(s)K(s))^{-1}G(s)K(s)\\
		R(s) &= K(s)(I + G(s)K(s))^{-1}
	\end{aligned}
\end{equation}
where $G(s)$ denotes the transfer function of a system, $K(s)$ denotes a controller. As shown in Fig. \ref{Figure2}, the sensitivity function is a transfer function from the disturbance $w$ and the reference signal $r$ to the tracking error $e$, which corresponds to the performance requirements of the time domain response. The complementary sensitivity function is a transfer function from the noise $n$ to the tracking error $e$, which can be interpreted as the weight of model uncertainty. The control sensitivity function is a transfer function from the disturbance $w$ to the control input $u$. To satisfy the design requirements and moderate the control actions as well as enhancing the robustness, the three weighting functions $W_{s}(s), W_{t}(s), W_{r}(s)$ are applied to the three closed-loop transfer functions $S(s), T(s), R(s)$, respectively. The weighting functions $W_{s}(s), W_{r}(s)$ have the following general forms:
\begin{equation}\label{equ.5}
	W_{s}(s) = \frac{s/M + \omega_{b}}{s + A \omega_{b}}
\end{equation}         
\begin{equation}\label{equ.6}
	W_{r}(s) = \frac{r_{max}/u_{max}s + \omega_{a}10^{-3}}{s + \omega_{a}}
\end{equation}  
where $M$ is the maximum value of the frequency response of the reciprocal of the weighting function, corresponding to the overshoot requirement for time-domain responses, $A$ is the maximum steady-state error requirement, $\omega_{b}$ is the designed bandwidth of a closed-loop system. $r_{max}$ is the upper limit of a reference signal, $u_{max}$ is the maximum control action that actuators can provide, $\omega_{a}$ is the bandwidth of actuators.

Further, the structured $H_{\infty}$ optimization problem can be defined as \cite{13}: to find controllers $K$, such that 
\begin{equation}\label{equ.7}
	\mathop{\rm min} \limits_{k \in K} \gamma	 
\end{equation} 
subject to: 
\begin{equation}\label{equ.8}
	\begin{Vmatrix} W_{s}(s)S(s) \\ W_{t}(s)T(s) \\ W_{r}(s)R(s) \end{Vmatrix}_{\infty} \leq \gamma
\end{equation}
where $K$ is the set of considered structured controllers.    

\subsubsection{Descriptions of the Perturbed Models}
During the transition flight, the airspeed of the dual-system UAV increases from 0 to 13 m/s (the stall speed of the UAV) continuously, which leads to varying transition dynamics. The varying dynamics during the transition poses a great challenge to the control design. The problem lies in the contradiction that transition dynamics are continuously varying, whereas controller design is usually based on discrete design points. To tackle this problem, a multi-model approach is adopted. That is, different discrete airspeed values are selected, each corresponding to a nominal design point. Accordingly, the nominal design models can be determined. For each design point, the slightly perturbed models with respect to the nominal model, which corresponds to the perturbed airspeed values around the nominal airspeed, can be characterized. Then, the control synthesis can be done mainly based on these nominal design points.      

To be specific, there are six nominal design points selected within the airspeed range from 0 to 13 m/s as listed in Table \ref{tab1}. For each design point, the uncertainty of the airspeed is considered by defining an upper bound and an lower bound around each nominal airspeed value. The variation ranges of the loss of control effectiveness $\gamma_{*}$ (* denotes L, M, N, T) are chosen as $[0,0.6]$. These defined variation ranges for the airspeed and the loss of control effectiveness are used for describing the uncertainties of the perturbed models with respect to the nominal ones. 

\begin{figure}
	\centering
	\includegraphics[width=0.7\linewidth]{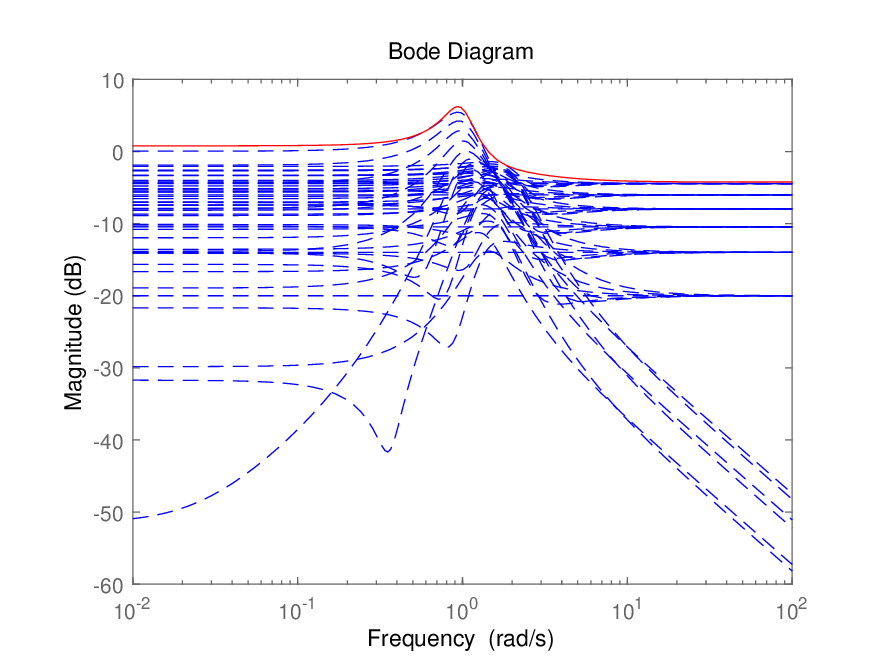}
	\caption{The multiplicative uncertainty weight, $\bar{V_{3}}$ = 4 m/s}
	\label{Figure3}
\end{figure}

In this research, multiplicative uncertainty descriptions are used to describe these model uncertainties, which are related to model uncertainty weights that are integrated into the mixed sensitivity optimization. For details on the definition of the uncertainty weight and the procedure to calculate it, please refer to \cite{17}. Following the procedure, the uncertainty weights of all the design points can be obtained. As an example, the uncertainty weight and relative errors of the design point No. 3 of the pitch dynamics model (nominal airspeed $\bar{V}$ = 4 m/s) are depicted in the frequency domain (as shown in Fig. \ref{Figure3}).  
\begin{table}
	\begin{center}
		\caption{The parameters of the nominal design points}
		\label{tab1}
		\begin{tabular}{ c  c  c  c}
			\hline \hline
			Design point NO.                    & \makecell{Nominal airspeed \\  $\bar{V}$ (m/s)}        & \makecell{Perturbed airspeed range \\ $[V_{\rm min},\ V_{\rm max}]$ (m/s) }     & \makecell{$\gamma_{*}$}\\
			\hline
			1                                   & 0                             &  [0, 0.8]                &  [0, 0.6]\\
			2                                   & 1                             &  [0.8, 2.5]              &  [0, 0.6]\\
			3                                   & 4                             &  [2.5, 5.5]              &  [0, 0.6]\\
			4                                   & 7                             &  [5.5, 8.5]              &  [0, 0.6]\\
			5                                   & 10                            &  [8.5, 11.5]             &  [0, 0.6]\\
			6                                   & 13                            &  [11.5, 13]              &  [0, 0.6]\\
			\hline  \hline         
		\end{tabular}
	\end{center}
\end{table}

\begin{figure}
	\centering
	\includegraphics[width=0.6\linewidth]{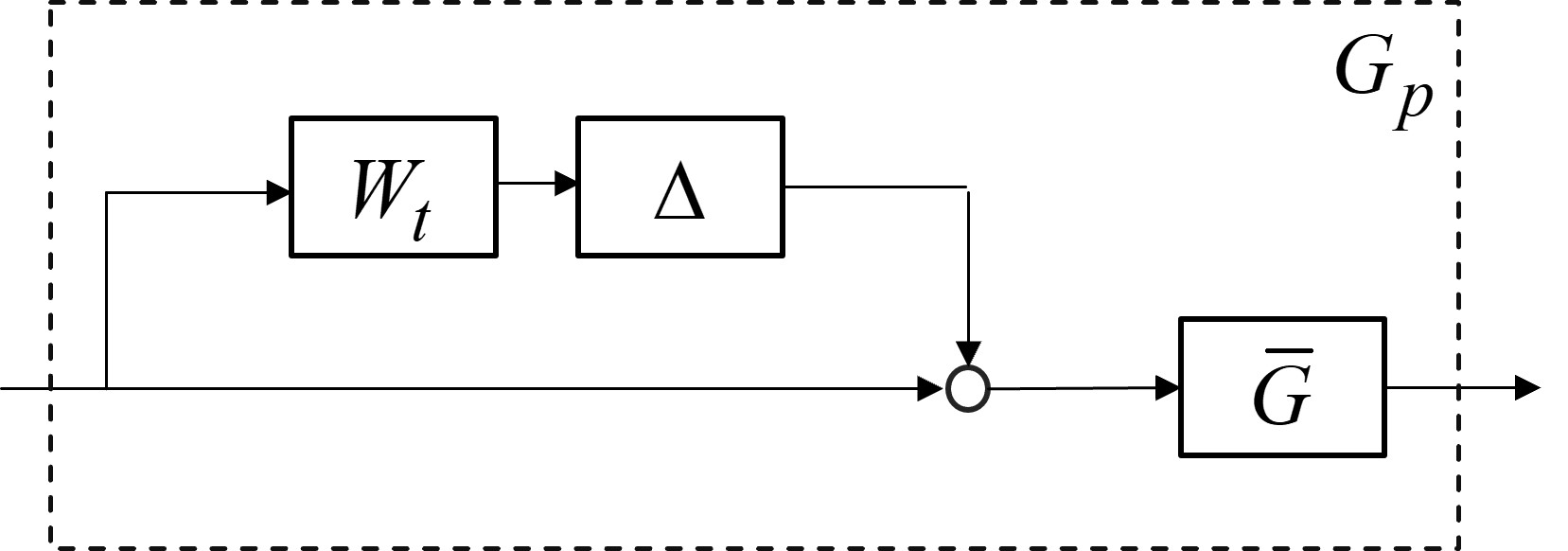}
	\caption{The perturbed model with multiplicative uncertainty}
	\label{Figure4}
\end{figure}

Therefore, the perturbed models can be denoted with the uncertainty weights and the nominal models within the framework of multiplicative uncertainty descriptions (as shown in Fig. \ref{Figure4}),
\begin{equation}\label{equ.11}
	G_{p,n}(s) = \bar{G}_{n}(s)(1 + W_{t,n}(s)\Delta(s))
\end{equation} 
with $ |\Delta(s)|_{\infty} < 1$.

\subsubsection{Control Synthesis}

\noindent 1. Altitude controller\\

In this paper, even though the attitude controllers are designed to be gain-scheduled, the altitude hold controller is still designed to be non-gain-scheduled with constant control parameters. The altitude hold controller for the transition flight used here is exactly the same as that in \cite{12}. The reader may refer to it for more details. \\

\noindent 2. Gain-scheduled fault-tolerant control design for the attitude \\
       
\begin{figure}
	\centering
	\includegraphics[width=0.7\linewidth]{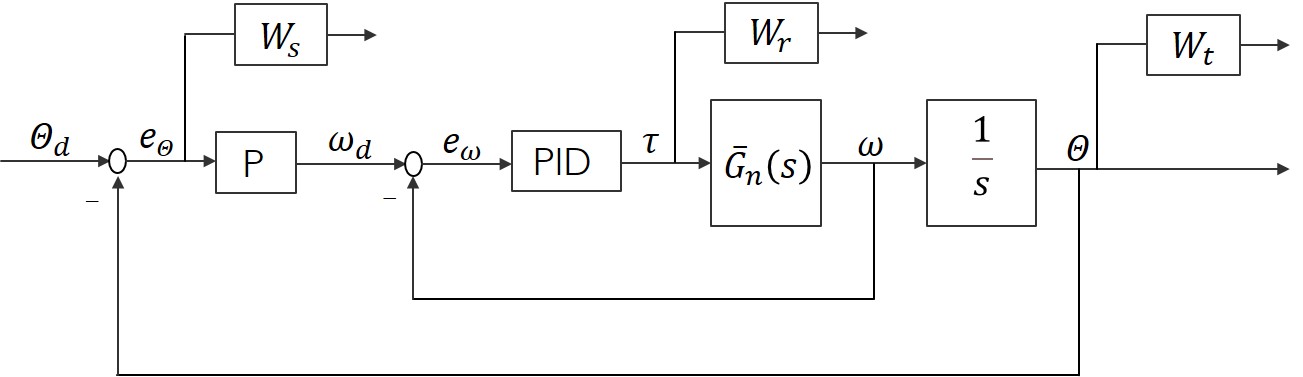}
	\caption{The control structure of the attitude controllers}
	\label{Figure5}
\end{figure}

Fault-tolerant controllers for the attitude are synthesized at the discrete design points using the structured $H_{\infty}$-based mixed sensitivity optimization. The attitude loops (including the roll, pitch and yaw loops) have the same control structure (as shown in Fig. \ref{Figure5}), which is a cascaded P-PID loop. To optimize the three closed-loop transfer functions $S(s), T(s), R(s)$, three weighting functions $W_{s}(s), W_{t}(s), W_{r}(s)$ are applied to the control error $e_{\theta}$, the control input $\tau$ and the control output $\theta$, respectively.     

Before performing the mixed sensitivity optimization, the three weighting functions need to be determined. The sensitivity weighting function $W_{s}(s)$ corresponds to performance requirements for time domain responses as described previously. The choice of parameters for $W_{s}(s)$ for all the design points is given in Table \ref{tab2}. The complementary sensitivity function $W_{t}(s)$ that can also be interpreted as a model uncertainty weight is already calculated in the previous subsection. For the control sensitivity function $W_{r}(s)$, the actuator bandwidths $\omega_{a}$ are chosen as 5 rad/s. Other parameters are determined according to their physical meaning as explained in Sec. \ref{Sec.3.1}. With the weighting functions determined, the optimization problem defined in Eq. \eqref{equ.8} can eventually be solved using the Matlab command 'systune'.  
\begin{table}
	\begin{center}
		\caption{The parameters of the determined sensitivity weighting functions}
		\label{tab2}
		\begin{tabular}{c  c  c  c  c  c  c  c  c  c}
			\hline \hline
			\multirow{2}{*}{Design point NO.}   & \multirow{2}{*}{$\bar{V}$ (m/s)}       &\multicolumn{2}{c}{Roll}   &                &\multicolumn{2}{c}{Pitch}  &     &\multicolumn{2}{c}{Yaw}\\
			\Xcline{3-4}{0.4pt}  \Xcline{6-7}{0.4pt}  \Xcline{9-10}{0.4pt}                                 
			&                       & $ M $            & $ \omega_{b} $ (rad/s) &                 & $ M$            & $ \omega_{b} $ (rad/s)    &                   & $ M $            & $ \omega_{b} $ (rad/s)\\
			
			\hline
			1           & 0          &2           &2.4       &   & 1.1      & 0.012   &    & 1.8      & 0.1\\
			2           & 1          &1.1         &0.5       &   & 2        & 0.2     &    & 1.3      & 0.002 \\
			3           & 4          &1.6         &1.1       &   &  15.3    & 0.059   &    & 1.8      & 0.003\\
			4           & 7          &2           &1.2       &   &  1.3     & 0.16    &    & 1.1      & 0.007  \\
			5           & 10         &1.9         &1.1       &   &  1.3    & 0.89     &    & 1.5      & 0.006   \\
			6           & 13         &1.8         &0.6       &   &  1.4     & 0.82    &    & 2.5      & 2.9     \\
			\hline \hline
		\end{tabular}
	\end{center}
\end{table}   

After the control synthesis, a set of discrete controllers at the design points is obtained. The gain-scheduled attitude controllers are realized by picewise linear interpolation of the control parameters between adjacent design points.   

\section{Robustness Analysis}\label{Sec.4}
In the previous section, the controllers for the attitude loops are synthesized at the discrete design points. However, the robustness including the robust stability and performance of the resulting controllers is still to be examined. In this section, the robustness analysis of the controllers is carried out. 
  
\subsection{Stability of the Nominal Systems}
\begin{figure}[htbp]  
	\centering
	\subfigure[Roll loop]
	{   
		\label{Figure6:sub1}
		\includegraphics[width=0.7\textwidth]{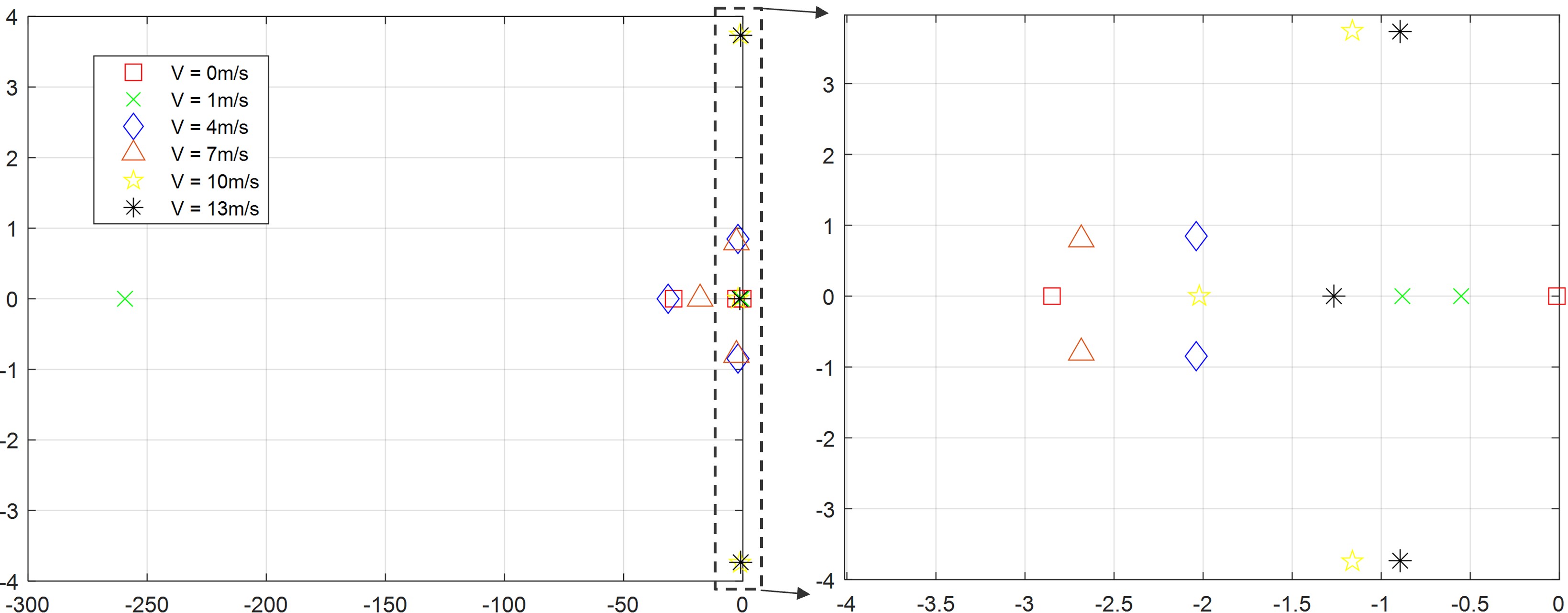}
	}
	\subfigure[Pitch loop]
	{   
		\label{Figure6:sub2}
		\includegraphics[width=0.7\textwidth]{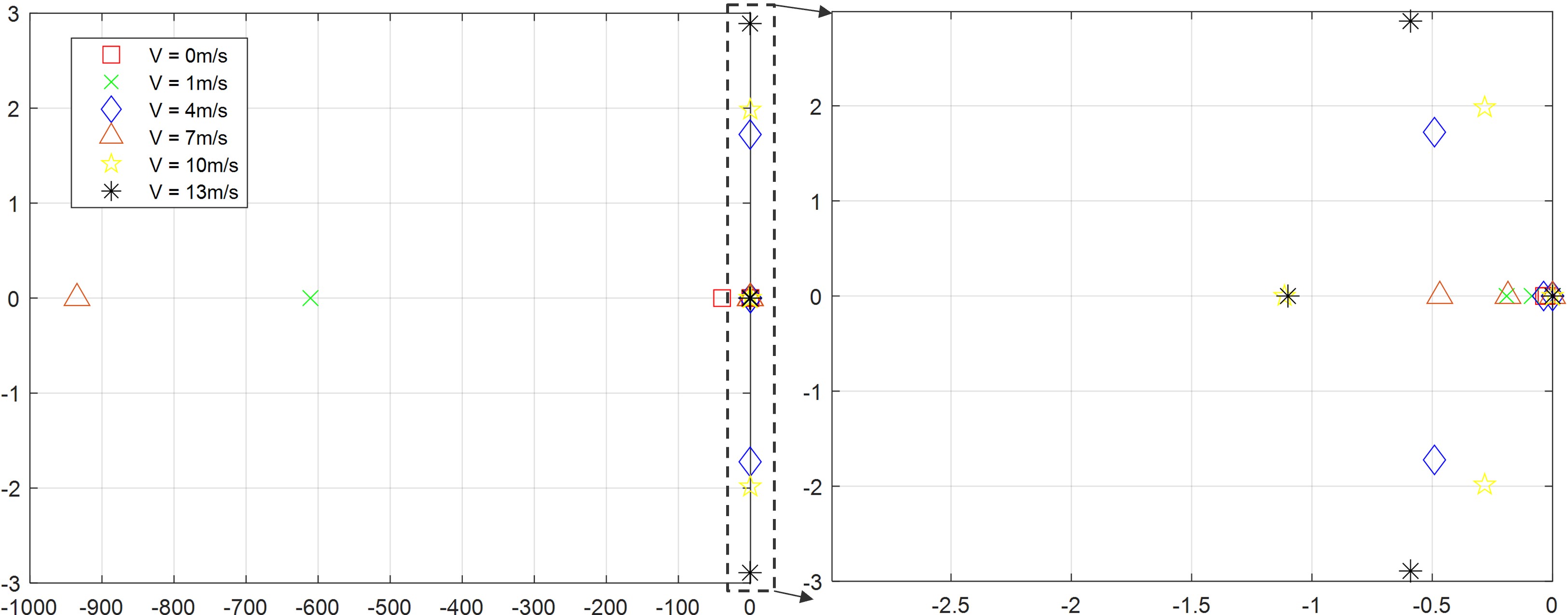}  
	}
	\subfigure[Yaw loop]
	{   
		\label{Figure6:sub3}
		\includegraphics[width=0.7\textwidth]{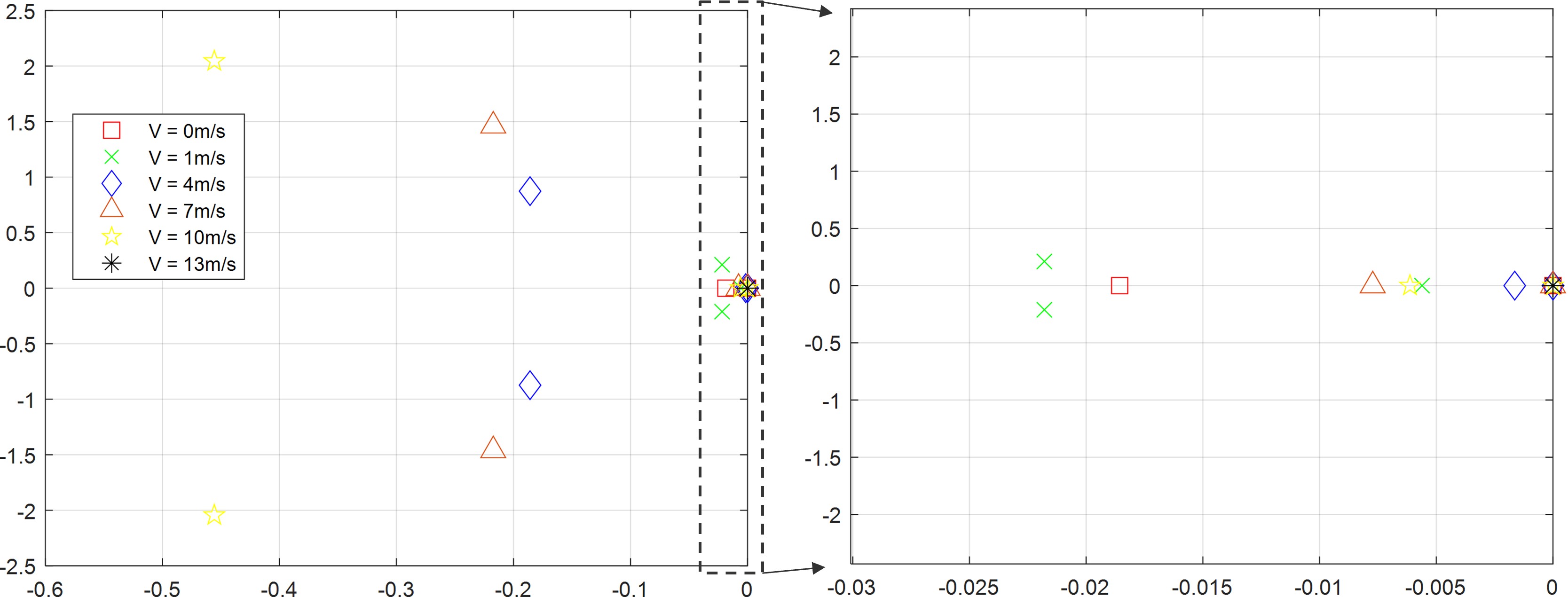}  
	}
	\caption{The poles of the closed-loop systems with nominal models}
	\label{Figure6}
\end{figure}

The discrete controllers for all the design points have already been obtained, and can now be taken as given for each design point. Nominal systems refer to control systems with controllers and nominal models closed in the loop. To ensure the robust stability and performance, the stability of the nominal systems must first be ensured. Therefore, the stability of the nominal systems at different design points is examined. For each attitude loop, the poles of the nominal closed-loop systems at the design points are shown in Fig. \ref{Figure6}. As can be seen, all the poles of the nominal closed-loop systems lie in the left half plane of the complex plane, indicating that the nominal closed-loop systems with the resulting controllers are stable at all the design points.

\subsection{Robust stability and performance}
The robust stability and performance analysis is based on the $\mu$-analysis, which is a powerful tool for analyzing the stability and performance of the uncertain closed-loop systems without the need to search through the whole uncertain model set.

The structured singular value (also known as $ \mu $) has the following definition \cite{13}:
\begin{equation}\label{equ.12}
	\mu(M) = \frac{1}{ \mathop{\rm min} \{ {k_{m}|det(I - k_{m} M \Delta) = 0} \} }
\end{equation}
where $ \Delta = \mathop{\rm diag} \{ \Delta_{i} \} $ denotes the structured uncertainty with $ \bar{\sigma}(\Delta) \leq 1 $. $ k_{m} $ is a scaling factor of the uncertainty $ \Delta $. $ M(s) $ is the transfer function from the output to the input of $ \Delta $. Here, please refer to \cite{13} for more details about the $ M-\Delta $ structure and its relationship with robust stability. The robust stability condition is thereby given as:

\begin{equation}\label{equ.13}
	\mu_{\Delta}(M(j\omega)) < 1.
\end{equation}

For the robust performance analysis, $ \mu $ can be further enhanced as $ \mu_{\hat{\Delta}} $. Where $ \hat{\Delta} = \mathop{\rm diag} \{ \Delta, \Delta_{P} \} $, $ \Delta_{P} $ is a fictitious block corresponding to the $ H_{\infty} $ performance requirements (refer to Chapter 8 in \cite{13} for more details). The sufficient and necessary condition for the robust performance is given by:
\begin{equation}\label{equ.14}
	\mu_{\hat{\Delta}}(N(j\omega)) < 1
\end{equation}
where $ N(s) $ is the transfer function from the output to the input of the augmented perturbations $ \hat{\Delta} $.

Then, the $\mu$-values of the robust stability and performance for all the design points are computed and given in Table \ref{tab3}. As shown in the table, for the design point No. 1 of the roll control, the design points No. 1 and 2 of the pitch control, the design points No. 1 and 2 of the yaw control, the $\mu$-values are greater than 1. This means the robust stability and performance conditions are violated for these design points, corresponding to hovering or near hovering conditions. The existing literature \cite{14,15} tells us that it is almost impossible to stabilize the UAV under these conditions. Nevertheless, since the UAV is in a hovering or near hovering condition, it can be easily controlled by a pilot if there is any unstable motion.  

As for all other design points, the $\mu$-values are strictly smaller than 1. According to the aforementioned robust stability and performance conditions, it is known to us that the robust stability and performance of these design points are ensured with the synthesized controllers under the uncertainties of both the aerodynamic coefficients resulting from the airspeed variation and the loss of control effectiveness caused by the actuator faults/failures. 

\begin{table}
	\begin{center}
		\caption{The structured singular values of the robust stability and performance}
		\label{tab3}
		\begin{tabular}{c  c  c  c  c  c  c  c  c}
			\hline \hline
			\multirow{2}{*}{Design point NO.}   &\multicolumn{2}{c}{Roll}   &                &\multicolumn{2}{c}{Pitch}  &     &\multicolumn{2}{c}{Yaw}\\
			\Xcline{2-3}{0.4pt}  \Xcline{5-6}{0.4pt}  \Xcline{8-9}{0.4pt}                                 
			& $ \mu_{\Delta} $            & $ \mu_{\hat{\Delta}} $ &                 & $ \mu_{\Delta} $            & $ \mu_{\hat{\Delta}} $    &                   & $ \mu_{\Delta} $            & $ \mu_{\hat{\Delta}} $ \\
			\hline
			1           & 0.996      & 1.051    &     &  2.75       & 2.748    &    & 3.57    & 3.71  \\
			2           & 0.872      & 0.953    &     &  1.963      & 1.964    &    & 0.306   & 1.086 \\
			3           & 0.758      & 0.993    &     &  0.93       & 0.987    &    & 0.898   & 0.997 \\
			4           & 0.731      & 0.959    &     &  0.9767     & 0.988    &    & 0.134   & 0.973 \\
			5           & 0.690      & 0.952    &     &  0.6855     & 0.974    &    & 0.523   & 0.977 \\
			6           & 0.639      & 0.831    &     &  0.69       & 0.984    &    & 0.859   & 0.988 \\
			\hline \hline           
		\end{tabular}
	\end{center}
\end{table}  

\section{Simulation Results}\label{Sec.5}
In the previous sections , the discrete controllers at all design points are synthesized. Using the gain-scheduling technique, the discrete controllers are able to work continuously. In this section, the obtained controllers are validated on a nonlinear simulator by assuming two different fault scenarios. For each fault scenario, the proposed GS SHIF FTC method is compared to the SHIF and LQR FTC approaches. In addition, the reference tracking performance of the three methods is quantitatively evaluated, which demonstrates the validity and advantages of the FTC method proposed in this paper.

\subsection{Case1: 50$\%$ partial loss of rotor 2b, 20$\%$ partial loss of elevator, ailerons, rudders}
In this section, partial loss of a single rotor and aerodynamic control surfaces is assumed for the nonlinear simulation. At the beginning of the simulation, the UAV is in hovering state at 20 s. Then the UAV starts transition flight and the airspeed starts to increase. At 22 s, 50$\%$ partial loss of rotor 2b and 20$\%$ partial loss of elevator, ailerons, rudders are injected into the system. The UAV continues the transition flight under the faulty condition. At almost 27.3 s, the airspeed reaches the stall speed of the aircraft (i.e., 13 m/s). After that, the UAV enters the fixed-wing mode.

Three FTC methods are applied and compared, including the linear quadratic regulator (LQR) method, the structured $H_{\infty}$ (SHIF) method and the gain-scheduled structured $H_{\infty}$ (GS SHIF) method. The LQR method is proven to be an effective FTC design method for actuator faults/failures of a quadrotor \cite{16}. In this paper, the selection of the LQR weights is similar to that in \cite{16}, but not exactly the same considering that the configurations of the dual-system UAV and the quadrotor are essentially different. The Q matrices are the same. And the R matrices are selected to be a diagonal matrix with diagonal elements equal to 0.5 instead of an identity matrix. As the number of the rotors is doubled, the control moments that the rotors can provide are also doubled. Accordingly, the weights for control efforts become half of those for the quadrotor case. The SHIF FTC method for the dual-system UAV is reported in \cite{12}, in which a structured $H_{\infty}$ controller with constant control parameters is designed for the FTC of dual-system UAV transition flight. 

\begin{figure}
	\centering
	\includegraphics[width=0.9\linewidth]{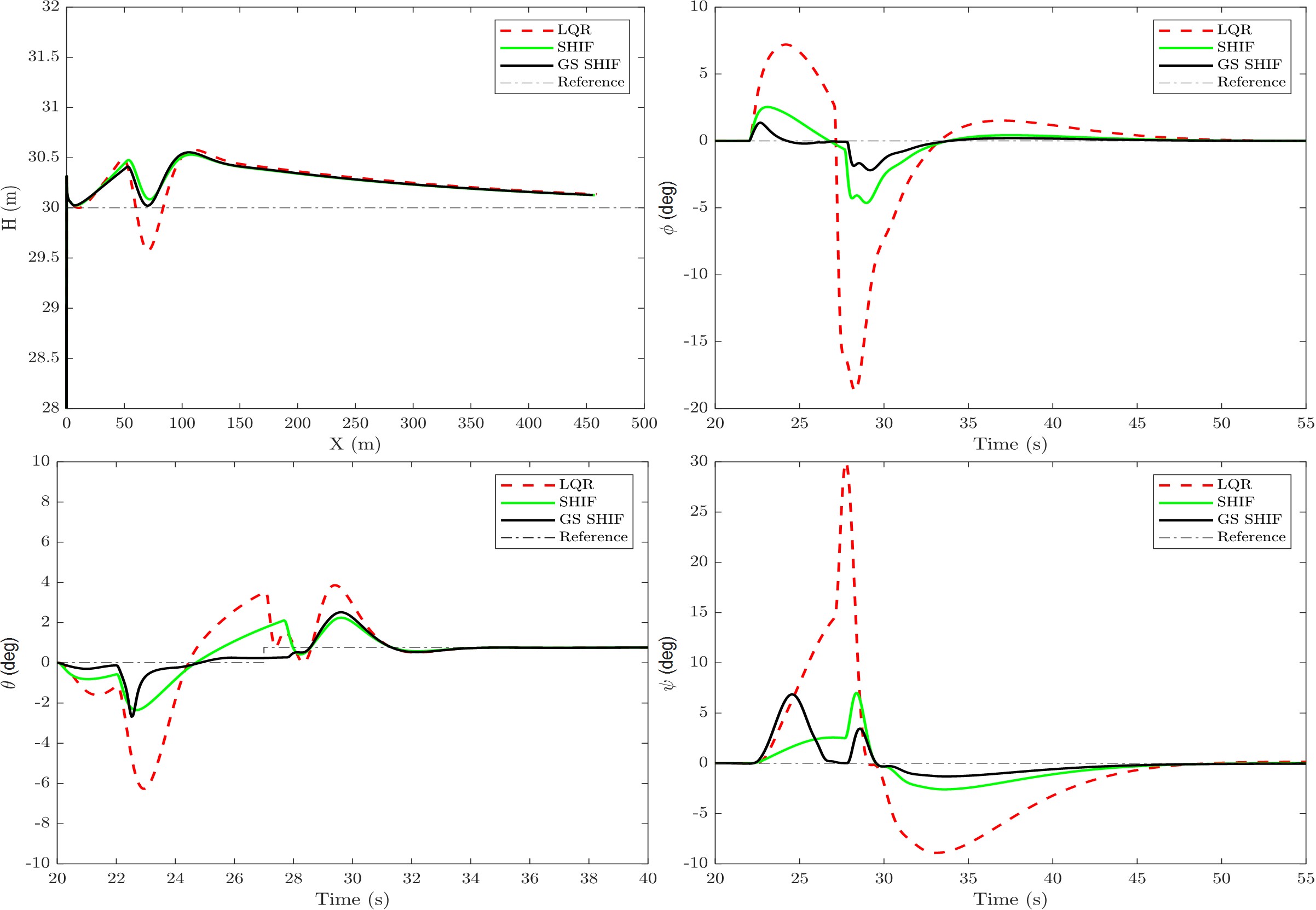}
	\caption{The longitudinal trajectories and attitude variations for Case 1}
	\label{Figure7}
\end{figure}

The FTC systems designed using the three methods are each simulated with the nonlinear simulator under the same fault scenario, respectively. The longitudinal trajectories and the time histories of the attitude angles for the three methods are shown in Fig. \ref{Figure7}. As can be seen, the plots of the altitude in the longitudinal plane for the three methods almost coincide before 22 s, corresponding to the no-fault case. After 22 s, even with the actuator faults, the altitudes all converge consistently to 30 m. As for the attitude angles, after fault occurrence, there are large variations in the attitude angles. But with the FTC systems, the attitude angles eventually converge to the steady state values of 0, 0.7 and 0 degrees, respectively. The attitude variations for the GS SHIF control are smaller than those for the LQR and SHIF methods.  

\begin{figure}
	\centering
	\includegraphics[width=0.5\linewidth]{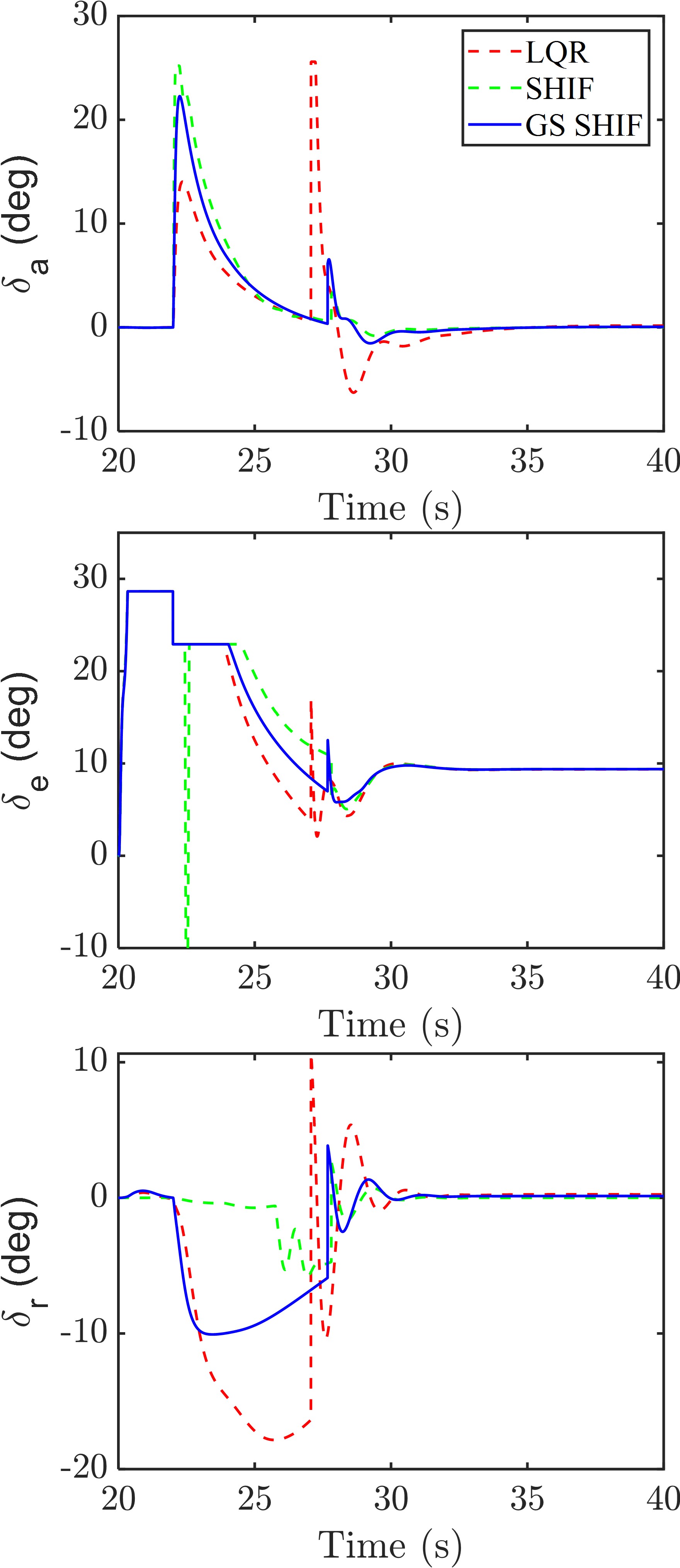}
	\caption{The time histories of the aerodynamic control surfaces deflections for Case 1}
	\label{Figure8}
\end{figure}

\begin{figure}
	\centering
	\includegraphics[width=0.7\linewidth]{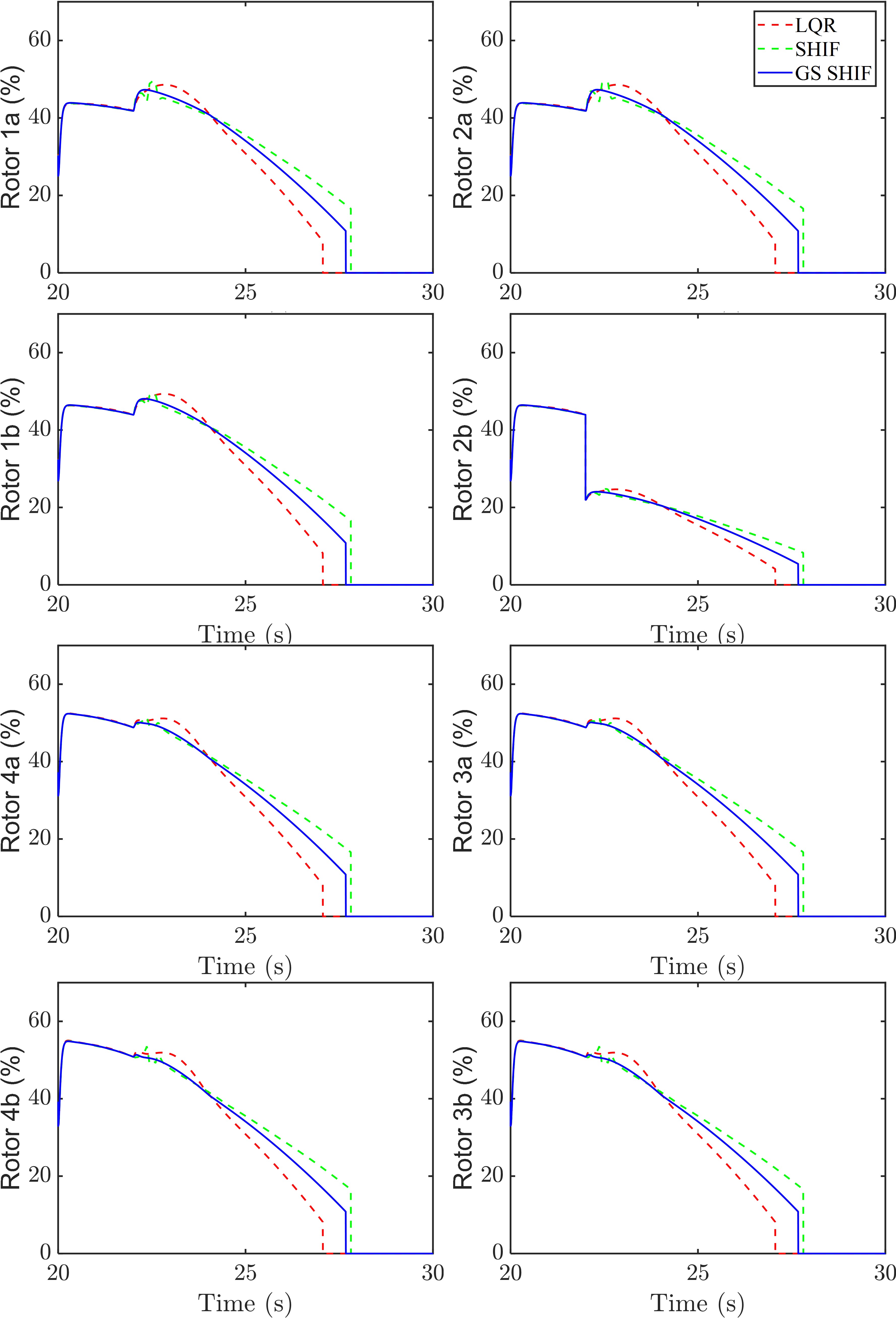}
	\caption{The time histories of the throttle percentages of the vertical rotors for Case 1}
	\label{Figure9}
\end{figure}

The time histories of the actuators including the aerodynamic control surfaces and the vertical rotors are shown in Figs. \ref{Figure8} and \ref{Figure9}, respectively. The time histories of the actuators for the three methods are quite similar to each other. From Fig. \ref{Figure8}, it can be seen that, after fault occurrence, the ailerons and rudders are activated to counteract the unbalanced torques caused by the rotor faults. The elevator deflection becomes 80$\%$ of the value before 22 s (corresponding to the no-fault case). The elevator deflection finally converges to a constant value of 9.38 degrees that is required for a cruise flight in the fixed-wing mode. For the throttle percentages of the vertical rotors, in Fig. \ref{Figure9}, the throttle percentage of rotor 2b becomes one half of the value before 22 s. Simultaneously, almost all the throttle percentages of the other rotors are increased to compensate for the loss of the force and moments caused by the actuator faults.             

\subsection{Case2: 50$\%$ partial loss of rotors 1a, 2b, 4a, 20$\%$ partial loss of elevator, ailerons and rudders}
In this subsection, to illustrate the capability of the developed FTC system to tolerate more complicated fault scenarios, simultaneous partial loss of rotors and aerodynamic control surfaces is further considered, i.e., 50$\%$ partial loss of rotors 1a, 2b, 4a, 20$\%$ partial loss of elevator, ailerons and rudders. The simulation setup remains unchanged. Again, the FTC systems with the three control methods are simulated under this new fault scenario.

\begin{figure}
	\centering
	\includegraphics[width=0.9\linewidth]{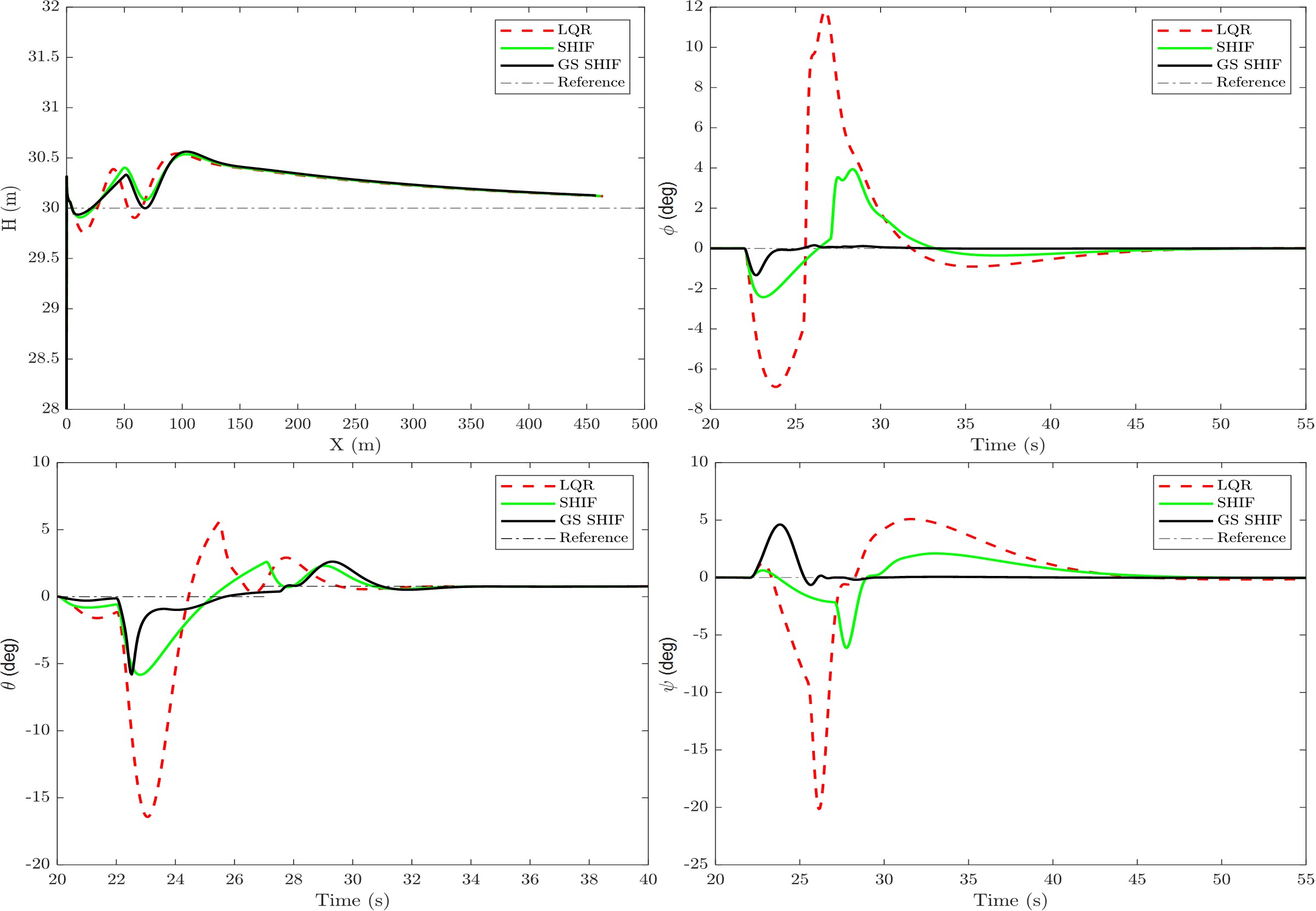}
	\caption{The longitudinal trajectories and attitude variations for Case 2}
	\label{Figure10}
\end{figure}

The longitudinal trajectories and the time histories of the attitude angles are shown in Fig. \ref{Figure10}. The longitudinal trajectories are similar to those of Case 1. For the attitude variations, after fault occurrence, there are significant increases in the attitude angles. The attitude variations for the LQR method are much larger than those for the other two methods. Finally, the FTC systems corresponding to the three methods are able to make the attitude angles converge to the steady state values.   

\begin{figure}
	\centering
	\includegraphics[width=0.5\linewidth]{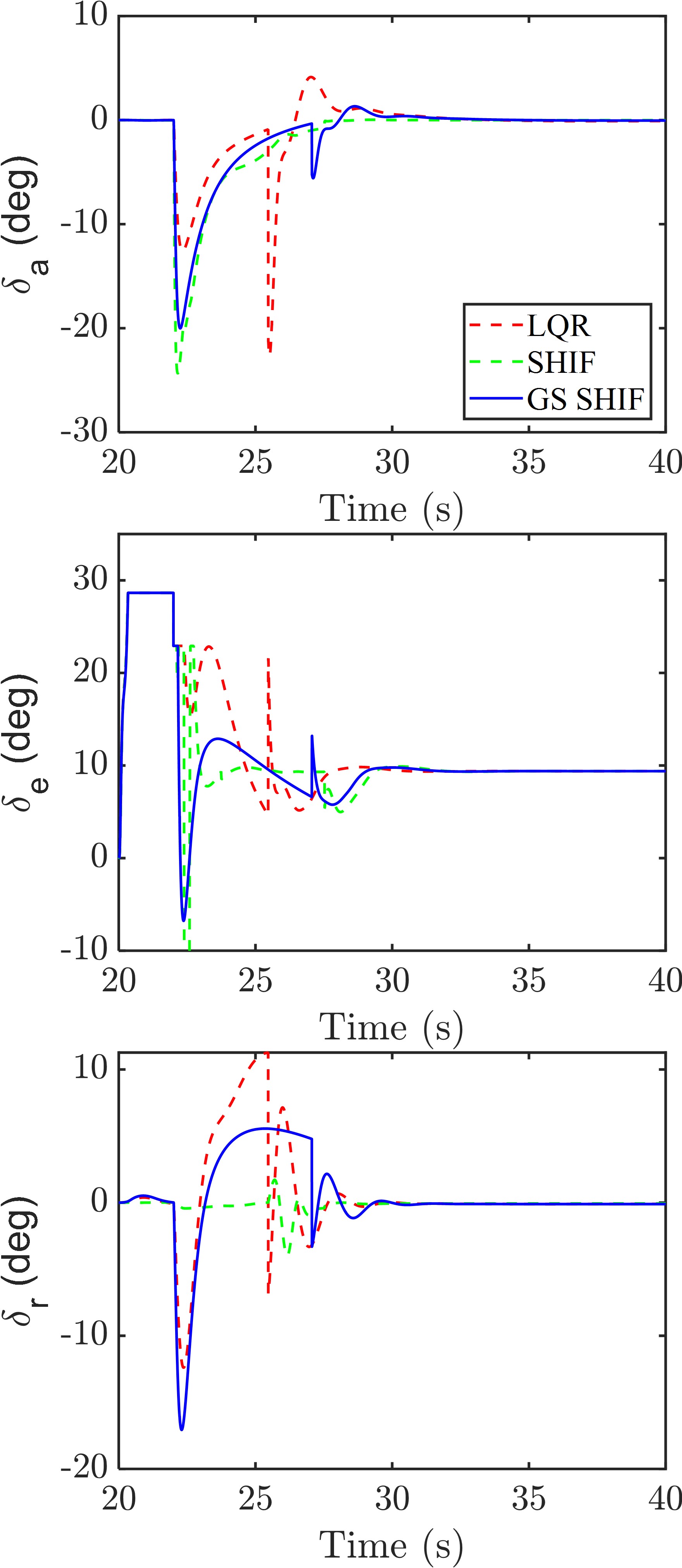}
	\caption{The time histories of the aerodynamic control surfaces deflections for Case 2}
	\label{Figure11}
\end{figure}

\begin{figure}
	\centering
	\includegraphics[width=0.7\linewidth]{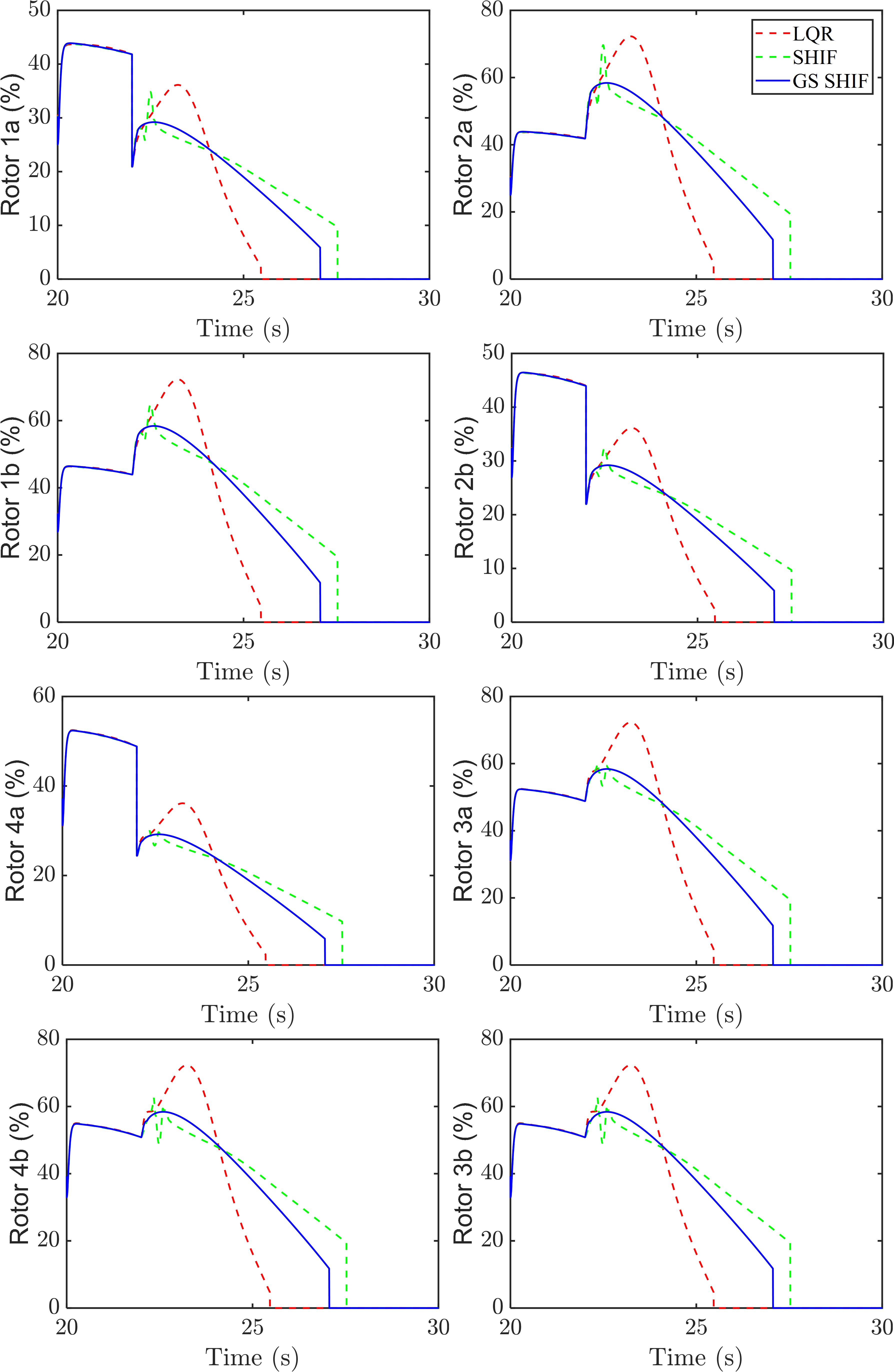}
	\caption{The time histories of the throttle percentages of the vertical rotors for Case 2}
	\label{Figure12}
\end{figure}

The time histories of the actuators of Case 2 are shown in Figs. \ref{Figure11} and \ref{Figure12}. Figure \ref{Figure11} shows the time histories of the aerodynamic control surface deflections. Before 22 s, the aileron and rudder deflections are close to zero. And the elevator deflection is at the saturated value. After fault occurrence, the ailerons and rudders are deflected to balance the additional rolling and yawing moments resulting from the actuator faults. For the elevator, the deflection becomes 80$\%$ of the saturated value due to the partial loss fault. And the elevator deflection deviates immediately from the saturated value in order to control the pitch angle instead of staying at the saturated value for a while in Case 1. The time histories of the throttle percentages of the vertical rotors are shown in Fig. \ref{Figure12}. As can be seen, after 22 s, the throttle percentages of rotors 1a, 2b, 4a are reduced by half due to the partial loss fault. Other than these three rotors, the throttle percentages of the other rotors are increased immediately to different extents compared with the values before 22 s. Because of the simultaneous partial loss of these three rotors, the resulting relatively large loss of force and moments is compensated by the throttle percentage increases in the other rotors. In both Case 1 and Case 2, the different FTC methods result in different transition flight durations.              
 
\subsection{Evaluation of reference tracking performance}       
\begin{figure}
	\centering
	\includegraphics[width=0.5\linewidth]{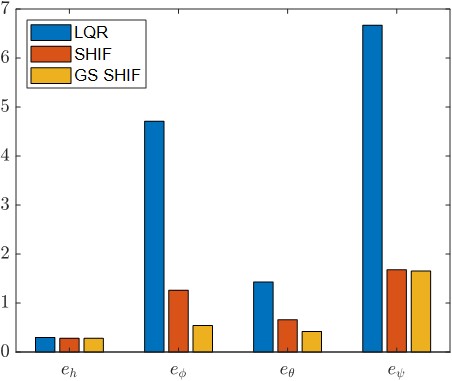}   
	\caption{The RMSE of the reference tracking for Case 1}
	\label{Figure13}
\end{figure}

\begin{figure}
	\centering
	\includegraphics[width=0.5\linewidth]{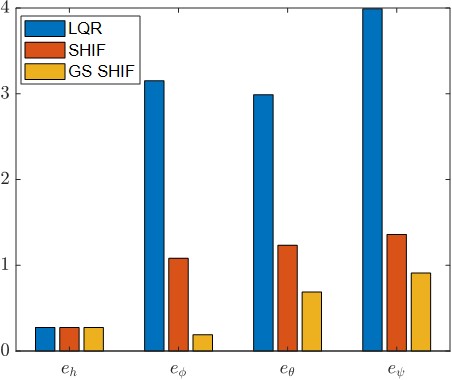}
	\caption{The RMSE of the reference tracking for Case 2}
	\label{Figure14}
\end{figure}

To quantitatively evaluate the reference tracking performance of the FTC methods, the root mean squared error (RMSE) of the attitude tracking is calculated for each of the three methods. The RMSEs for both Case 1 and Case 2 are given in Table \ref{tab4}. The corresponding bar graphs are shown in Figs. \ref{Figure13} and \ref{Figure14}, respectively. As can be seen, the altitude tracking errors for the three methods are almost equal in both Case 1 and Case 2, because the altitude controllers are all designed to have constant control parameters instead of being designed as gain-scheduled. As for the attitude tracking errors, in Case 1 (see Fig. \ref{Figure13}), the roll and pitch angle tracking errors for the GS SHIF method are significantly smaller than those for the LQR and SHIF methods. The yaw angle tracking error for the GS SHIF method is almost equal to that for the SHIF method, but far smaller than that for the LQR method. In Case 2 (see Fig. \ref{Figure14}), the tracking errors of all the attitude angles for the GS SHIF control are smaller than those for the LQR and SHIF controls. Therefore, in general, the GS SHIF FTC method provides significantly better attitude tracking performance than the LQR and SHIF FTC methods for the dual-system UAV transition flight under the considered actuator faults.

\begin{table}
	\begin{center}
		\caption{Root mean squared errors of the reference tracking for Case 1 and Case 2}
		\label{tab4}
		\begin{tabular}{ c  c  c  c  c  c}
			\hline \hline
			Case                  &FTC Method                 & $e_{h}$ (m)         & $e_{\phi}$ (deg)       & $e_{\theta}$ (deg)      & $e_{\psi}$ (deg) \\
			\hline
			\multirow{3}{*}{1}    & LQR              & 0.2997           & 4.7117           & 1.432             & 6.6718    \\
			                      & SHIF             & 0.2817           & 1.2638           & 0.6627            & 1.683     \\
			                      & GS SHIF          & 0.2812           & 0.5411           & 0.4223            & 1.6551    \\
			\hline
			\multirow{3}{*}{2}    & LQR              & 0.2761           & 3.1536           & 2.9905            & 3.9939    \\
			                      & SHIF             & 0.2744           & 1.0816           & 1.2335            & 1.3610    \\
			                      & GS SHIF          & 0.2758           & 0.1921           & 0.6891            & 0.9096    \\
			\hline  \hline         
		\end{tabular}
	\end{center}
\end{table}

\section{Conclusion}\label{Sec.6}
This paper presents a novel GS SHIF PFTC design for the transition flight of a dual-system UAV. Based on the previously proposed airspeed-dependent linearized attitude dynamics models, the loss of control effectiveness is further introduced into the linearized models. These new design models are used to synthesize the FTC system based on the multi-model approach, in which the model uncertainties arising from both the aerodynamic coefficient uncertainties caused by airspeed variation and the loss of control effectiveness induced by actuator faults/failures are represented by multiplicative uncertainty descriptions. The uncertainty descriptions are integrated into the structured $H_{\infty}$ synthesis for each design point. The robust stability and performance of the resulting discrete controllers at the design points are examined. The analysis results show that the obtained controllers satisfy the robustness requirements. Eventually, the developed GS SHIF PFTC system is validated on the nonlinear simulator by assuming two different fault scenarios of partial loss of the vertical rotors and the aerodynamic control surfaces. To illustrate the superiority of the proposed FTC method, the FTC systems designed with the three methods are simulated and compared. The conclusions of this paper are as follows:
\begin{enumerate}
	\item The proposed GS SHIF FTC design method is feasible for the dual-system UAV transition flight. This method allows us to consider multiple model uncertainties apart from the loss of control effectiveness caused by actuator faults/failures. As a general approach, it is applicable to the FTC design of many other uncertain systems with multiple parametric uncertainties.
	\item Within the framework of the robust control theory, the use of multiplicative uncertainty descriptions and a multi-model approach for the FTC design can usually reduce the design effort and improve the efficiency of the control system development compared to the existing self-scheduled structured $H_{\infty}$ approach.
	\item The simulation results show that the developed GS SHIF PFTC system with constant control parameters is able to deal with simultaneous partial loss of the vertical rotors and the aerodynamic control surfaces. The comparative simulation results suggest that the proposed GS SHIF FTC method can provide better reference tracking performance than the LQR and SHIF FTC methods, demonstrating the effectiveness and superiority of the proposed method.
\end{enumerate}
               
\section*{Acknowledgments}
The first author acknowledges the financial support from China Scholarship Council (CSC, No.202006020041).

\bibliography{sample}

\end{document}